\documentclass[sigconf]{acmart}

\usepackage{booktabs} 
\usepackage{tabularx} 
\usepackage{caption}
\usepackage{subcaption}
\usepackage[utf8]{inputenc}
\usepackage{listings}
\usepackage{mathtools,xparse}
\usepackage{enumitem}
\usepackage{float}

\usepackage{balance}
\usepackage{csquotes}

\usepackage{tikz-network}
 
\copyrightyear{2018}
\acmYear{2018}
\setcopyright{rightsretained}
\acmConference[MIS2]{WSDM workshop on Misinformation and Misbehavior Mining on the Web}{2018}{Marina Del Rey, CA, USA}
\acmDOI{10.475/123_4}
\acmISBN{123-4567-24-567/08/06}

\acmPrice{}

\begin{document}

\title{``Like Sheep Among Wolves'':   \\ Characterizing Hateful Users on Twitter}

\author{Manoel Horta Ribeiro, Pedro H. Calais,  Yuri A. Santos, Virgílio A. F. Almeida, Wagner Meira Jr.}
\orcid{1234-5678-9012}
\affiliation{%
  \institution{Universidade Federal de Minas Gerais}
  \city{Belo Horizonte} 
  \state{Minas Gerais} 
  \country{Brazil}
}
\email{{manoelribeiro, pcalais, yurisantos, virgilio, meira}@dcc.ufmg.br}
\renewcommand{\shortauthors}{Manoel Horta Ribeiro. et al.}


\begin{abstract}
Hateful speech in Online Social Networks (OSNs) is a key challenge for companies and governments, as it impacts users and advertisers, and as several countries have strict legislation against the practice. 
This has motivated work on detecting and characterizing the phenomenon in tweets, social media posts and comments. 
However, these approaches face several shortcomings due to the noisiness of OSN data, the sparsity of the phenomenon, and the subjectivity of the definition of hate speech.
This works presents a user-centric view of hate speech, paving the way for better detection methods and understanding.
We collect a Twitter dataset of $100,386$ users along with up to $200$ tweets from their timelines with a random-walk-based crawler on the retweet graph, and select a subsample of $4,972$ to be manually annotated as hateful or not through crowdsourcing. 
We examine the difference between user activity patterns, the content disseminated between hateful and normal users, and network centrality measurements in the sampled graph.
Our results show that hateful users have more recent account creation dates, and more statuses, and followees per day. 
Additionally, they favorite more tweets, tweet in shorter intervals and are more central in the retweet network, contradicting the ``lone wolf'' stereotype often associated with such behavior.
Hateful users are more negative, more profane, and use less words associated with topics such as hate, terrorism, violence and anger.
We also identify similarities between hateful/normal users and their 1-neighborhood, suggesting strong homophily.
\end{abstract}

\ccsdesc[500]{Human-centered computing~Social network analysis}
\ccsdesc[500]{Human-centered computing~Empirical studies in collaborative and social computing}

\keywords{hate speech, online social networks, hateful users}

\maketitle

\section{Introduction}
\label{sec:introduction}
Hate speech can be defined as \textit{"language that is used to express hatred towards a targeted group or is intended to be derogatory, to humiliate, or to insult the members of the group"}~\cite{davidson2017automated}. 
The importance of understanding the phenomenon in Online Social Networks (OSNs) is manifold.
For example, countries such as Germany have strict legislation against the practice~\cite{stein1986history}, the presence of such content may pose problems for advertisers~\cite{youtubeboycott} and users~\cite{sabatini2017online}, and manually inspecting all possibly hateful content in OSNs is unfeasible~\cite{schmidt2017survey}.
Furthermore, the blurry line between banning such behavior from platforms and censoring dissenting opinions is a major societal issue~\cite{rainie2017future}.

This scenario has motivated a body of work that attempts to characterize and automatically detect such content~\cite{djuric2015hate, kwok2013locate, burnap2016us,magu2017detecting,warner2012detecting,schmidt2017survey}. These create representations for tweets, posts or comments in an OSN, \textit{e.g.}~word2vec~\cite{mikolov2013efficient}, and then classify content as hateful or not, often drawing insights on the nature of hateful speech on the granularity level of tweets or comments. 
However, in OSNs, the meaning of such content is often not self-contained, referring, for instance, to some event which just happened, and 
the texts are packed with informal language, spelling errors, special characters and sarcasm~\cite{dhingra2016tweet2vec,riloff2013sarcasm}.  Furthermore, hate speech itself is highly subjective, reliant on temporal, social and historical context, and occurs sparsely~\cite{schmidt2017survey}.
These problems, although observed, remain largely unaddressed~\cite{davidson2017automated,magu2017detecting}.  

Fortunately, the data in posts, tweets or messages, are not the only signals we may use to study hate speech in OSNs.
Most often, these signals are linked to a profile representing a person or institution.
Characterizing and detecting hateful \emph{users} shares much of the benefits of detecting hateful content and presents plenty of opportunities to explore a richer feature space.
Twitter's guideline for hateful conduct captures this intuition, stating that \textit{some Tweets may seem to be abusive when viewed in isolation, but may not be when viewed in the context of a larger conversation}~\cite{twitterguidelines}.

Analyzing hateful \textit{users} rather than \textit{content} is also attractive because other dimensions may be explored, such as the user's activity and connections in the network.
For example, in \textit{Twitter}, it is possible to see the number of tweets, followers, and favorites a user has.
It is also possible to extract influence links among users who retweet each other, analyzing them in a larger network of influences. 
This allows us to use network-based metrics, such as betweenness centrality~\cite{freeman1977set} and also to analyze the neighborhood of such users.
Noticeably, although several studies characterize and detect hateful speech in text~\cite{davidson2017automated,waseem2016hateful}, no study that the authors are aware of focuses explicitly on the dimension of hateful \textit{users} in OSNs.

\newpage

\begin{figure}[h]
\centering
\includegraphics[width=0.75\linewidth]{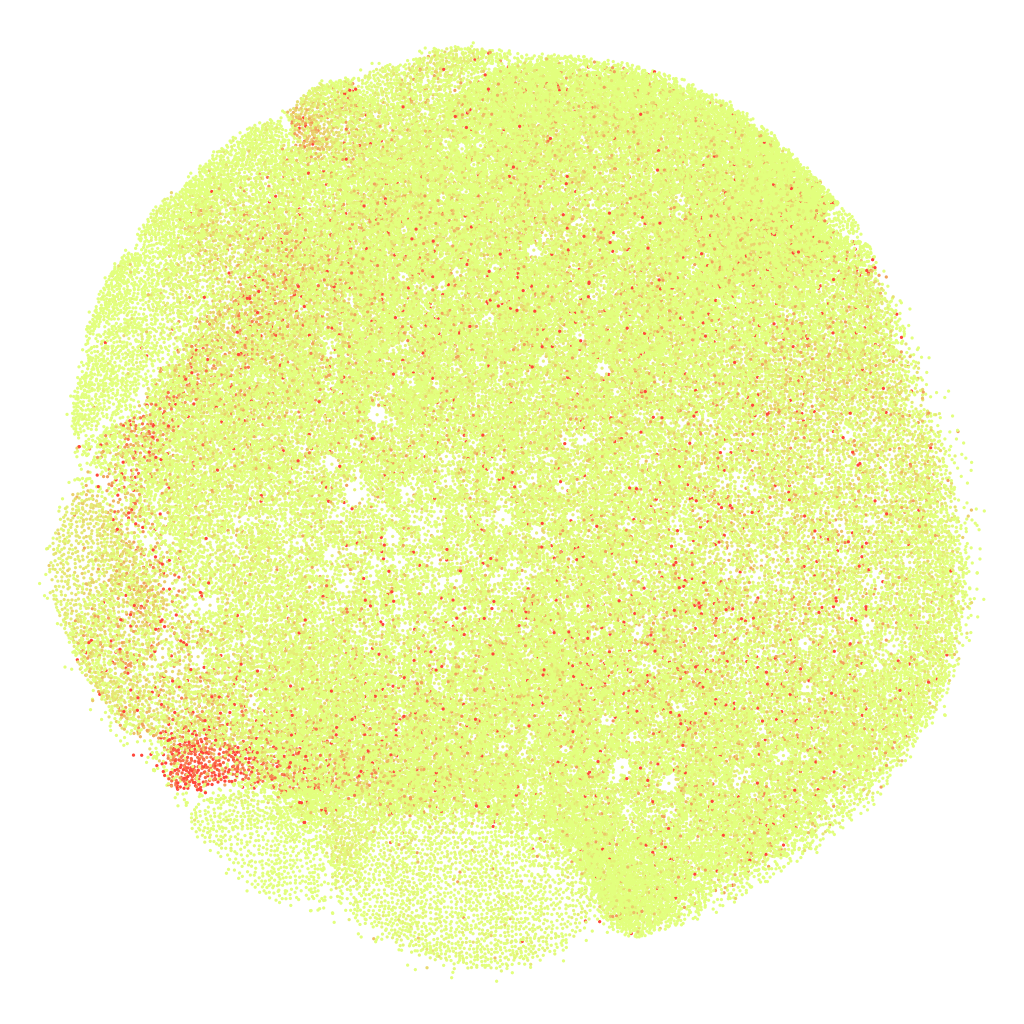}
\caption{Network of $100,386$ users sampled from Twitter after our diffusion process. Red nodes indicate the proximity of users to those who employed words in our lexicon.}
\label{fig:hintwitter}
\end{figure}

In this paper we focus on identifying and characterizing hateful users on Twitter, which we define in accordance with Twitter's hateful conduct guidelines~\cite{twitterguidelines}.
We collect a dataset of $100,386$ users along with up to $200$ tweets from their timelines with a random-walk-based crawler on Twitter's retweet-induced graph.
We identify users that employed a set of hate speech related words, and generate a subsample selecting users that are in different ``distances" to these to be manually annotated as hateful or not through crowdsourcing. This is explained in Section~\ref{sec:data_col}. 
We create a dataset containing $4,972$ manually annotated users, of which $544$ were labeled as hateful.
We ask the following research questions:

\vspace{0.4mm}

\noindent
\textit{\textbf{Q1}: Are the attributes of and the content associated with hateful users different from normal ones?}

\noindent
\textit{\textbf{Q2}: How are hateful users characterized in terms of their global position in the network and their local neighborhood of interactions?}

\vspace{0.5mm}

\noindent
To address these questions, we perform experiments in our collected dataset. We \textit{(i)}~examine attributes provided by Twitter's API, such as number of followers and creation date as well as attributes related to users activity; \textit{(ii)}~perform a sentiment and lexical analysis on the content present in each user's timeline; and \textit{(iii)}~compare centrality measures such as betweenness and eigenvector centrality between hateful and normal users. We also examine these statistics for users in the 1-neighborhood on the retweet graph.

Our results show that hateful users tweet more and within smaller intervals, and favorite other tweets significantly more than the normal ones. They also are more negative according to lexicon-based sentiment analysis and use more swear words.
Hateful users have follow more people per day than normal ones, and use vocabulary related to categories such as hate, anger, shame, violence and terrorism \textit{less} frequently. Also, the median hateful user have higher network centrality according to several metrics, contradicting the "lone wolf" behavior often associated with the practice~\cite{lonewolf}. This analysis held similar results when we looked at the 1-neighborhood of hateful and normal users. Our code is available online~\footnote{https://github.com/manoelhortaribeiro/AbusiveUsersOSNs}.
%
\section{Definitions}
\label{sec:definitions}
\noindent
\paragraph{Retweet-Induced Graph}
We define the retweet-induced graph $G$ as a directed graph $G=(V,E)$ where each node $u \in V$ represents a user in Twitter, and each edge $(u_1,u_2)\in E$ represents a retweet in the network, where the user $u_1$ has retweeted user $u_2$. Retweet-graphs have been largely used in the social network analysis, with previous work suggesting that retweets are better than  followers to judge the influence of users~\cite{cha2010measuring}. Notice that influence flows in the opposite direction of retweets, and thus we actually work on the graph with inverted edges. Intuitively, given that a lot of people retweet $u_i$ and $u_i$ retweets nobody, $u_i$ may still be a central and influential node.

\noindent
\paragraph{Hateful Users}
Defining which users are hateful is non-trivial as it derives from the definition of hateful speech, which is not widely agreed upon~\cite{sellars2016defining}.
We choose to define hateful users in accordance to Twitter's hateful conduct guidelines, which state users \textit{may not promote violence against or directly attack or threaten other people on the basis of race, ethnicity, national origin, sexual orientation, gender, gender identity, religious affiliation, age, disability, or disease. We also do not allow accounts whose primary purpose is inciting harm towards others on the basis of these categories}~\cite{twitterguidelines}.

\noindent
\paragraph{Offensive Language}
Other concept we employ is that of offensive language, which has been shown to be correlated with hateful content~\cite{davidson2017automated}. While there doesn't exist a universal definition of offensive language, we employ Waseem et. al definition of explicit abusive language, which defines it as \textit{language that is unambiguous in its potential to be abusive, for example language that contains racial or homophobic slurs}~\cite{waseem2017understanding}. Importantly, the use of this kind of language does not necessarily imply hate speech.
%
\section{Data Collection}
\label{sec:data_col}
\begin{figure*}[h]
\centering
\includegraphics[width=1\textwidth]{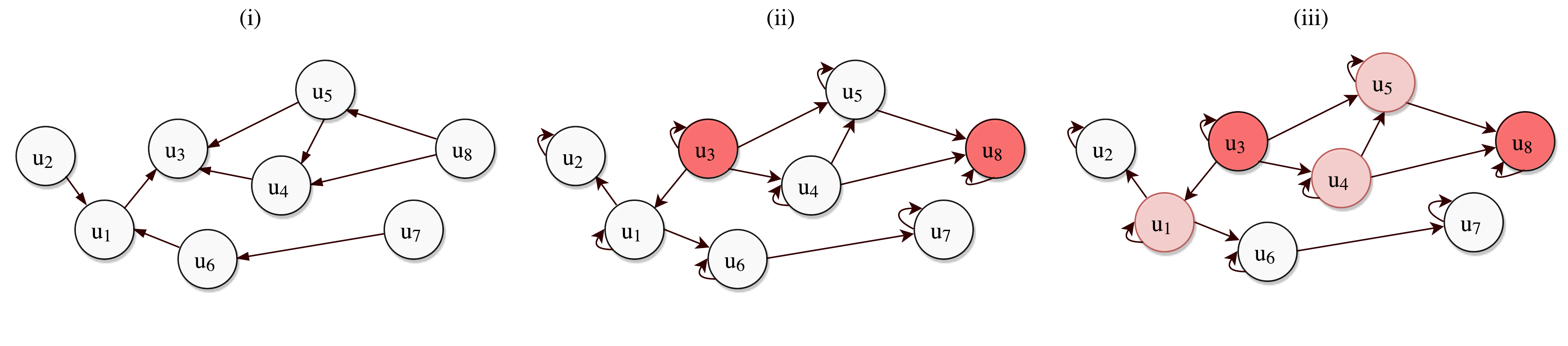}
\caption{Depiction of our diffusion process. \textit{(i)}~We begin with graph $G$ from the retweet-induced graph we sample from twitter, \textit{(ii)} We revert the direction of the edges (as it is the way influence flows), add self loops to every node, and mark the users who employed one of the words in our lexicon, \textit{(iii)} We iteractively update the belief of other nodes.}
\label{fig:all_users_nets}
\end{figure*}

Most existing work that detects hate speech on Twitter employ a lexicon-based data collection, which involves sampling only tweets that contain certain words~\cite{davidson2017automated, waseem2016hateful, burnap2016us, magu2017detecting}, such as \texttt{wetb*cks} of \texttt{fagg*t}. As we are trying to characterize hateful users, it would not be appropriate to rely solely on this technique, as we would get a sample heavily biased towards users who used these words. Furthermore this methodology presents problems even for dealing with the problem strictly on a tweet-based level. Some examples are:
\begin{itemize}
\item Statements may subtly disseminate hate with no offensive words, as in the sentence \texttt{"Who convinced Muslim girls they were pretty?"}~\cite{davidson2017automated, schmidt2017survey, waseem2016hateful}.
\item Hate groups may employ code words that are apparently benign, such as  \texttt{"skypes"}, to reference minorities demeaningly, creating a truly adversarial setting~\cite{magu2017detecting, operationgoogle}.
\end{itemize}

\noindent
Thus, we employ a more elaborate data collection process, which involves collecting a sample of Twitter's English speaking users, selecting a subsample of these users to be annotated as hateful or not hateful, and, finally, annotating them using a crowdsourcing service. These are described in the upcoming paragraphs.

\paragraph{Sampling Twitter.} As we do not have access to the full Twitter graph, we are faced with the challenge of obtaining a representative sample of it. 
Although there are several ways which users relate to each other in Twitter, we choose the retweet graph, in accordance with existing literature~\cite{cha2010measuring}.
Sampling the retweet-induced is hard as we can only observe out-coming edges, or in other words, given a user's timeline, we can obtain all users he or she retweeted, but not all users who retweeted them (due to API limitations). 
Furthermore, it is known that any unbiased in-degree estimation is impossible without sampling most of these ``hidden'' edges in the graph~\cite{ribeiro2012sampling}.
Acknowledging these limitations, we employ Ribeiro et al. Direct Unbiased Random Walk (\textit{DURW}), algorithm, which constructs an undirected graph in real time and estimates out-degrees distribution efficiently by occasionally jumping to a random node in the undirected graph~\cite{ribeiro2010estimating}. Fortunately, however, in the retweet graph the outcoming edges of a user represent the other users they (usually~\cite{guerra2017antagonism}) endorse.
With this strategy, we collect $100,386$ users and $2,286,592$ retweet edges along with the $200$ most recent tweets for each one of the users (including quotes, retweets and replies).

\paragraph{Selecting a Subsample to Annotate} After sampling Twitter, we are faced with the problem of selecting the subset of the data which will be annotated as hateful or not. If we choose the users uniformly at random, we risk having a very insignificant percentage of hate speech in the subsample. On the other hand, if we choose only users that use obvious hate speech related features, such as offensive racial slurs, we will bias our sample with only tweets with this language. In this case, for example, we would not capture code-words as the ones mentioned in Magu et. al~\cite{magu2017detecting}. We:
\begin{enumerate}
\item Create a lexicon of words that are mostly used in the context of hate speech. This is unlike other work~\cite{davidson2017automated}, as we don't consider words that are employed in a hateful context but often used in the everyday life in a harmless way (\textit{e.g.} \texttt{n*gger});
\item Run a diffusion process on the graph based on DeGroot's Learning Model~\cite{golub2010naive}, assigning a initial belief $p_{i}^{(0)} = 1$ to each user $u_i$ who employed the words in the lexicon;
\item Divide the users in $4$ strata according to their associated beliefs after the diffusion process, and perform a stratified sampling, obtaining up to $1500$ user per strata.
\end{enumerate}

\noindent
We create our lexicon with words from Hatebase.org~\cite{hatebase}, and ADL's hate symbol database~\cite{adlhate}. We choose words such as \texttt{holohoax}, \texttt{racial treason} and \texttt{white genocide} as they are less likely to be used in a non-hateful context. Furthermore, as we run the diffusion process later, we do not risk having a sample which is excessively small or biased towards some vocabulary.
Notice that the difference here is that we use the lexicon as a starting point to select regions of the graph to be sampled.

We briefly present our diffusion model, as illustrated in Figure~\ref{fig:all_users_nets}. Let $A$ be the adjacency matrix of our retweeted induced graph $G=(V,E)$ where each node $u \in V$ represents a user and each edge $(u,v)\in E$ represents a retweet. We have that $A(u,v) = 1$ if $u$ retweeted $v$. 
We create a transition matrix $T$ by inverting the edges in $A$ (as the influence flows from the retweeted user to the user who retweeted him or her), adding a self loop to each of the nodes and then normalizing each row in $A$ so it sums to $1$. This means each user is equally influenced by every user he or she retweets.
We then associate a belief $p_{i}^{(0)} = 1$ to every user who employed one of the words in our lexicon, and  $p_{i}^{(0)} = 0$ to all who didn't. Lastly, we create new beliefs $\mathbf{p}^{(t)}$ using the updating rule:

\begin{equation}
\mathbf{p}^{(t)} = T\mathbf{p}^{(t-1)}
\end{equation}

\noindent
Notice that the all the beliefs converge $p_{i}^{(t)}$ to the same value as $t \rightarrow \infty$, thus we run the diffusion process with $t=2$. Notice also that $p_{i}^{(t)} \in [0,1]$. With this real value associated with each user, we get 4 strata by randomly selecting up to $1500$ users with $p_{i}$ in the intervals $[0,.25)$, $[.25,.50)$, $[.50,.75)$ and $[.75,1]$.

\paragraph{Annotating Hateful Users}
We annotate $4,972$ users as hateful or not using \textit{Crowdflower}, a crowdsourcing service. The annotators were given the definition of hateful conduct according to Twitter's guidelines, and asked to annotate each user with the question:

\begin{displayquote}
\textit{Does this account endorse content that is humiliating, derogatory or insulting towards some group of individuals (gender, religion, race, nationality) or support narratives associated with hate groups (white genocide, holocaust denial, jewish conspiracy, racial superiority)?}
\end{displayquote}

\noindent
Annotators were asked to consider the whole webpage context rather than only individual publications or isolate words, and given examples of terms and codewords in ADLs hate symbol database. Each user was independently annotated by $3$ annotators, and, if there was disagreement, he or she would be annotated by up to $5$ annotators. In the end the annotators identified $544$ hateful users.

\begin{figure*}[ht]
\centering
\includegraphics[width=.9\textwidth]{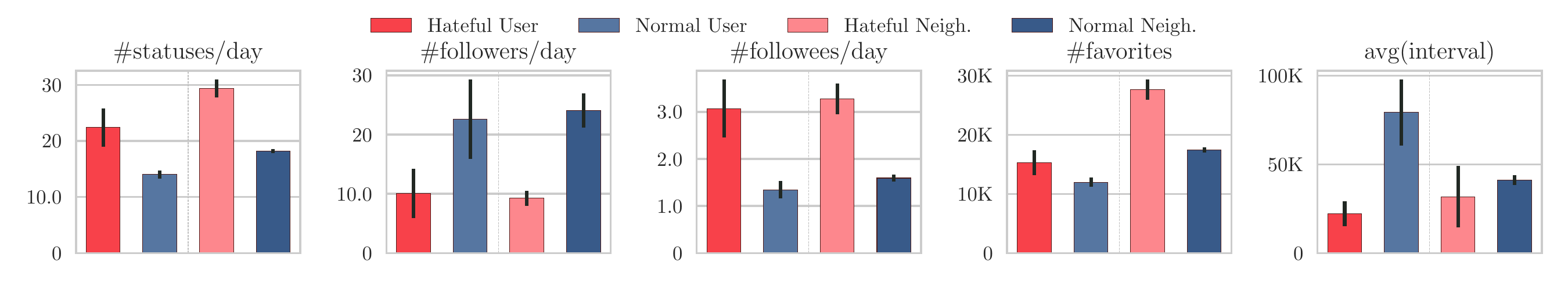}
\caption{Average values for several activity-related statistics for hateful users, normal users, and users in the neighborhood of those. \texttt{avg(interval)} was calculated on the $200$ tweets extracted for each user. Error bars represent $95\%$ confidence intervals. The legend used in this graph is kept in the remainder of the paper.}
\label{fig:attributes}
\end{figure*}

\begin{figure*}[ht]
    \centering
    \begin{minipage}{.48\textwidth}
        \centering
        \includegraphics[width=0.94\linewidth]{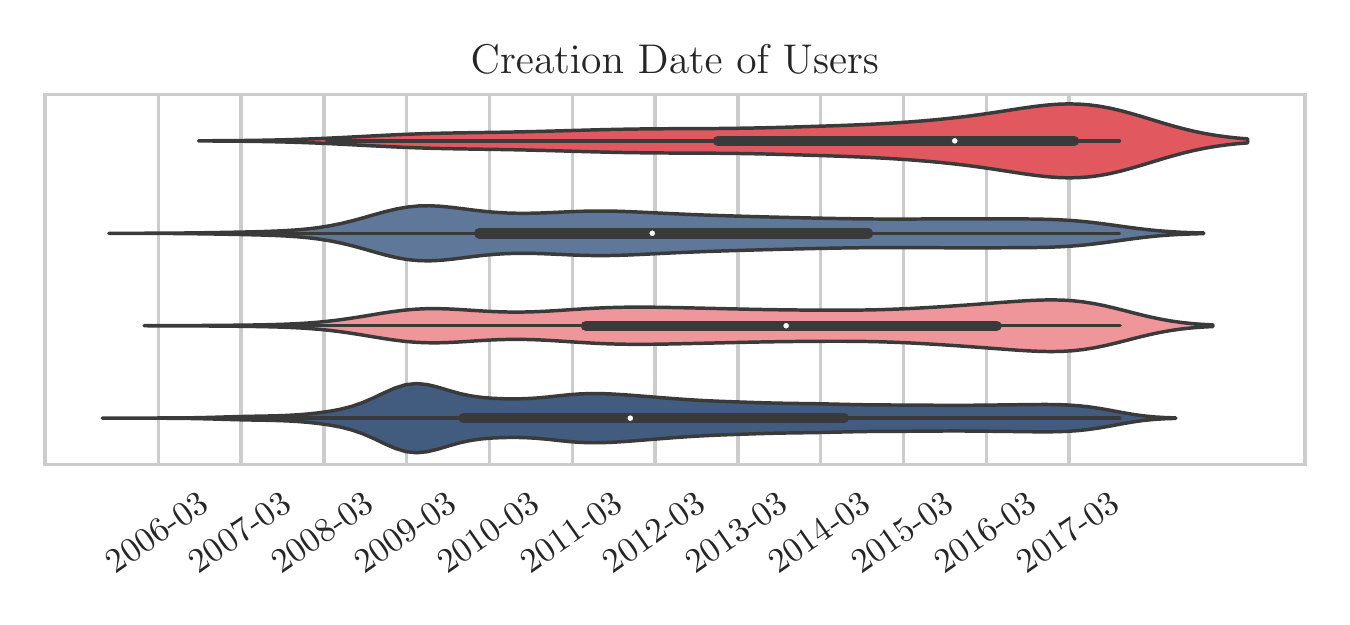}
        \caption{KDEs of the creation dates of user accounts. The white dot indicates the median and the thicker bar the first and third quartiles. Hateful users were created significantly later than their normal counterparts.}
        \label{fig:created_at}
    \end{minipage}\hspace{0.02\textwidth}
    \begin{minipage}{0.48\textwidth}
        \centering
        \includegraphics[width=0.90\linewidth]{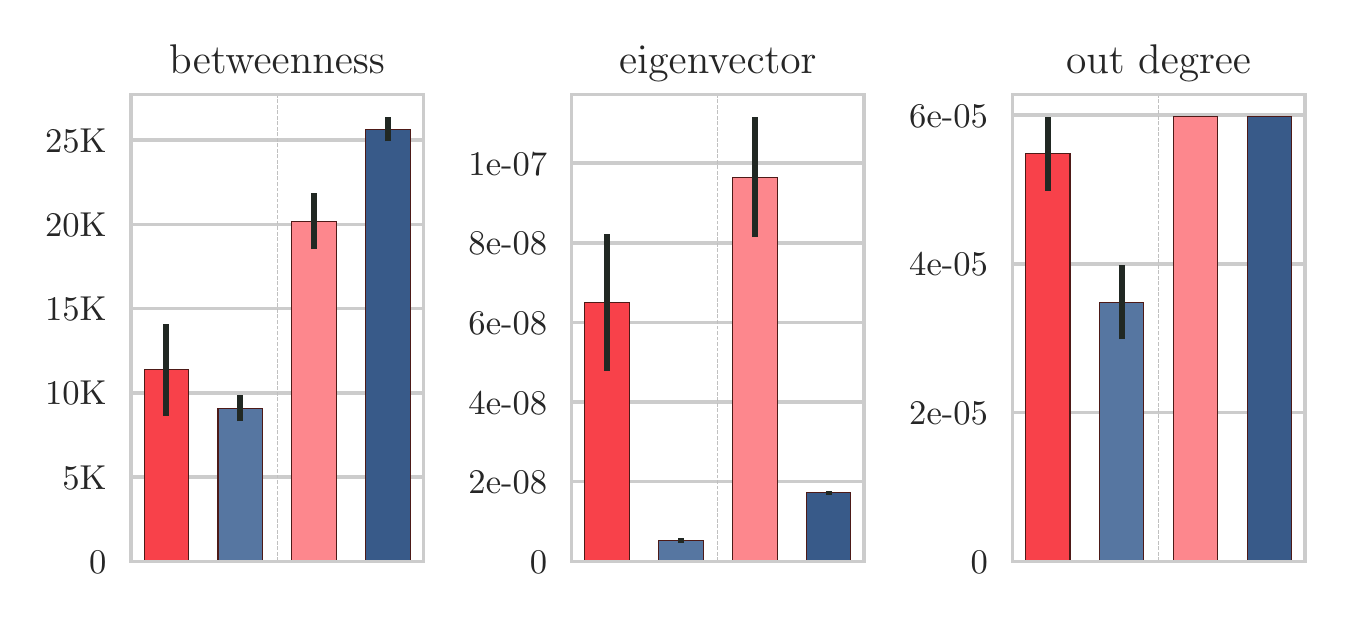}
        \caption{Median for network centrality metrics for hateful and normal users and their neighborhood calculated on the sampled retweet-induced graph.}
        \label{fig:betweenness}
    \end{minipage}
\end{figure*}
%
\section{Characterizing Hateful Users}
\label{sec:charac}

In this section we look at how hateful and normal users and their neighborhoods are different w.r.t. \textit{profile attributes}  provided by Twitter or inferred in the subgraph we sampled. Furthermore, we perform sentiment and lexical analysis on the \textit{content} produced.

\paragraph{Creation Dates} 
We begin by analyzing the account creation date of hateful and non-hateful users, as depicted in Figure~\ref{fig:created_at}. Notice that the hateful users were created later than the normal ones (p-value $< 0.001$). A hypothesis for this difference is that hateful users are banned more often than normal ones. This resonates with existing methods for detecting accounts created to sell followers, where methods using the distribution of creation date have been successful~\cite{viswanath2015strength}. We obtain similar results comparing the 1-neighborhood of such users, where the neighborhood of hateful users was also created more recently (p-value $< 0.001$).

\paragraph{User Activity} 
Other interesting metrics through which we can compare hateful and normal users, are the number of statuses, followers, followees and favorites a user has, and the interval in seconds between the tweets of each user. We show these statistics in Figure~\ref{fig:attributes}. We normalize the number of statuses, followers and followees by the number of days the users have since their account creation date. 
The results suggest that hateful users are "power users" in the sense that they tweet more, favorite more tweets by other people, and follow other users more (although they are less followed).
We also show these statistics to the users in the 1-neighborhood of hateful and normal users, which in practice represents the users these groups retweeted. The analysis yields similar results when we compare the 1-neighborhood of hateful and normal users: neighbors of hateful users have more statuses per day, more followees per day more favorites, but the difference on the interval between tweets is smaller.
It is hard to compare hateful/normal users and their neighborhood because of the distinct sampling methodology.

\paragraph{Network Centrality}  We also analyze different measures of centrality for the users and their neighborhood, as depicted in Figure~\ref{fig:betweenness}. The median hateful users and those in their neighborhood are more central in all measures when compared to their normal counterparts. This is an counter-intuitive finding, as hateful crimes, for example, have long been associated with ``lone wolves'', and anti-social people~\cite{lonewolf}. However, notice that, although the median for the centrality measurements for hateful user is bigger, the statistic for their average network centrality aren't. For example, none of the top $970$ most central users according to eigenvector centrality are hateful.

\begin{figure*}[ht]
\centering
\includegraphics[width=.95\textwidth]{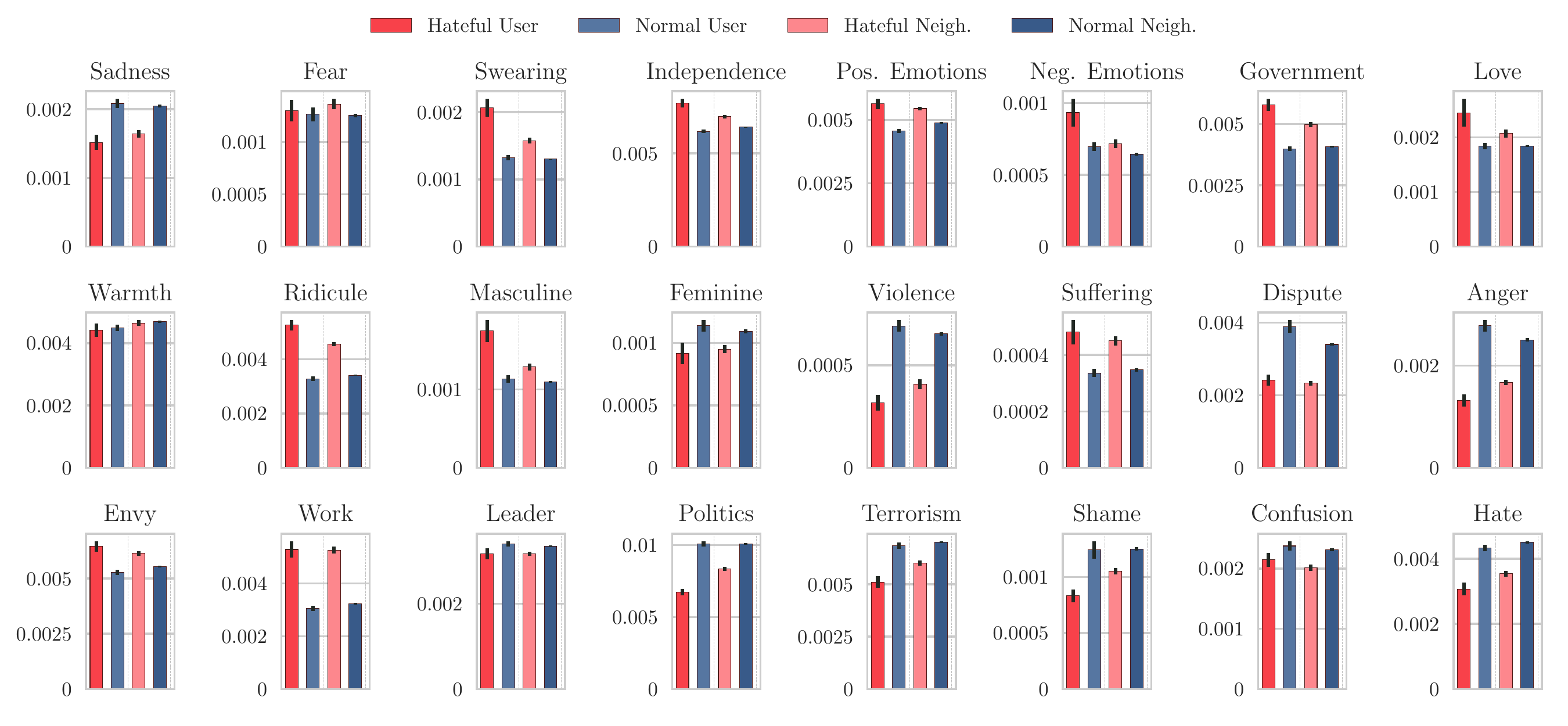}
\caption{Average values for the relative occurrence of several categories in \textit{Empath}. Notice that not all Empath categories were analyzed and that the to-be-analyzed categories were chosen before-hand to avoid spurious correlations. Error bars represent $95\%$ confidence intervals.}
        \label{fig:lex}
\end{figure*}

\begin{figure*}[ht]
    \centering
    \begin{minipage}{.48\textwidth}
        \centering
        \includegraphics[width=\textwidth]{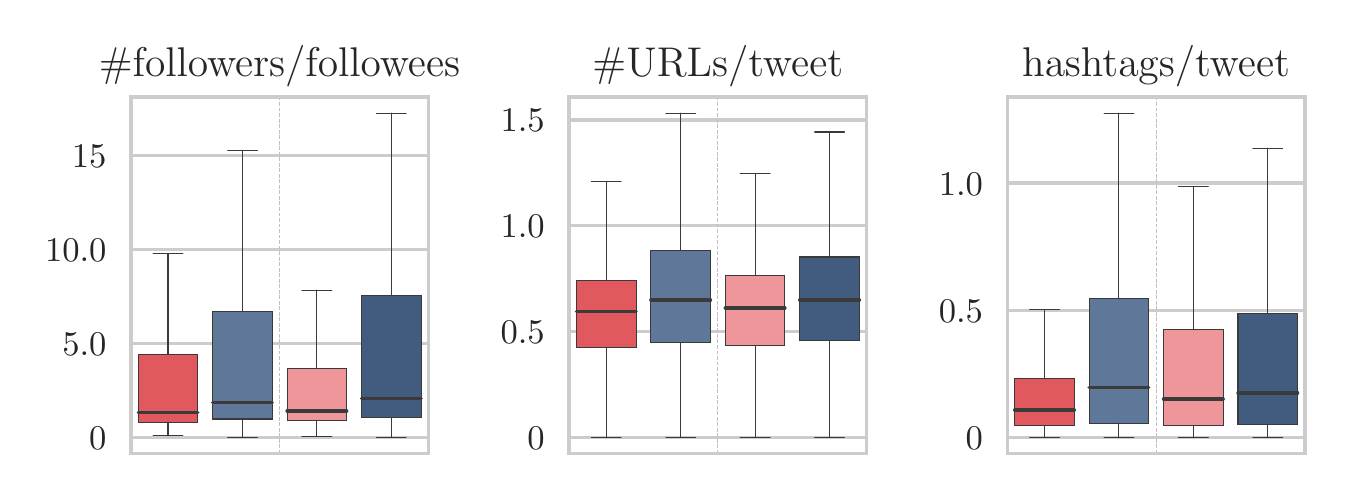}
        \caption{Boxplots for the distribution of metrics that indicate spammers. Hateful users and their neighborhood have slightly \textit{less} followers per followee, \textit{less} URLs per tweet, and \textit{less} hashtags per tweet.}
        \label{fig:spam}
    \end{minipage}\hspace{0.02\textwidth}
    \begin{minipage}{0.48\textwidth}
        \centering
        \includegraphics[width=\textwidth]{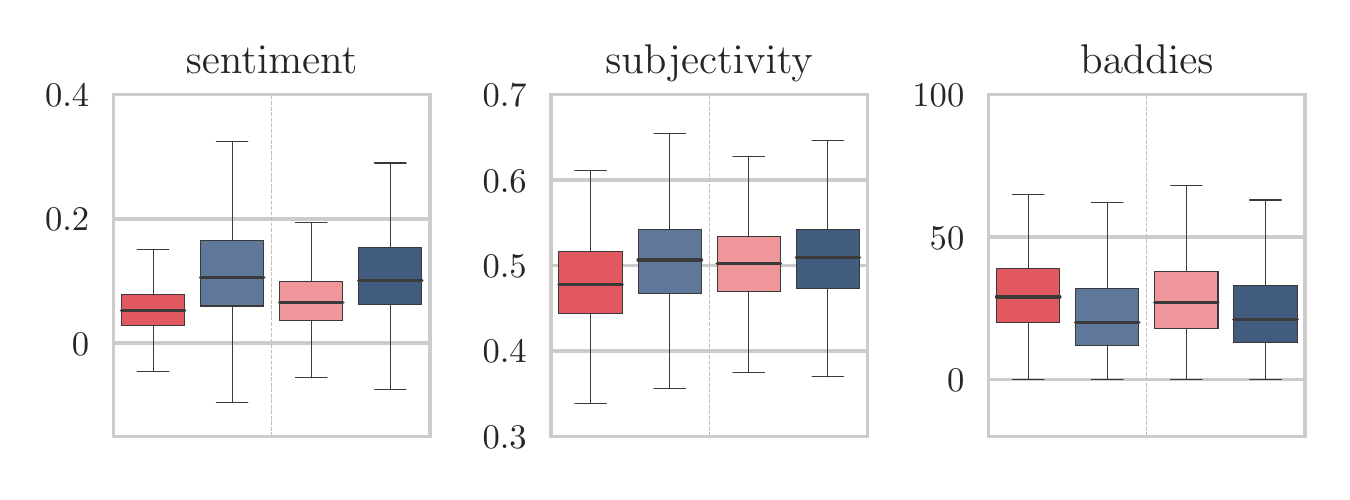}
        \caption{Boxplots for the distribution of sentiment, subjectivity and bad-words usage. Hateful users and their neighborhood are more negative, and use more bad words. Hateful users are less subjective.}
        \label{fig:sent}
    \end{minipage}
\end{figure*}

\paragraph{Spam}
It is interesting to consider the possible intersection between users that propagate hate speech and spammers, which have been widely studied.
First, it is worth to notice that our methodology of data collection is robust against spammers, as spammers often exploit trending topics or popular hashtags to post URLs. As our data collection don't specifically look for these trending hashtags or topics, it is intuitive that this problem is lessened significantly. To confirm this intuition, we analyze metrics that have been used by previous work to detect spammers, such as the number of URLs/tweet, and hashtags/tweet and the number of followers per followees~\cite{benevenuto2010detecting}. This boxplot of these distributions is shown on Figure~\ref{fig:spam}. We find that hateful users use, in average, \textit{less} hashtags (p-value $< 0.001$) and less URLs (p-value $< 0.001$) per tweet than normal users. The same analysis holds if we compare the 1-neighborhood of hateful and non-hateful users (also with p-values $< 0.001$). Additionally, we also find that in average normal users have more followers per followees than hateful ones (p-value $< 0.005$), which also happens for their neighborhood (p-value $< 0.001$). 
This suggests that the hateful users are not spammers, and thus were probably annotated as hateful or suspended for abusive behavior. Notice that it is not possible to extrapolate this finding to all hateful users in Twitter, as maybe there are other types that spread hate speech tagging messages in popular hashtags or trending topics.
Notice also that this doesn't necessarily mean that these accounts are not bots, although manual inspection by the authors suggests otherwise. 

\begin{figure*}[ht]
    \centering
    \begin{minipage}{.48\textwidth}
        \centering
        \includegraphics[width=0.96\linewidth]{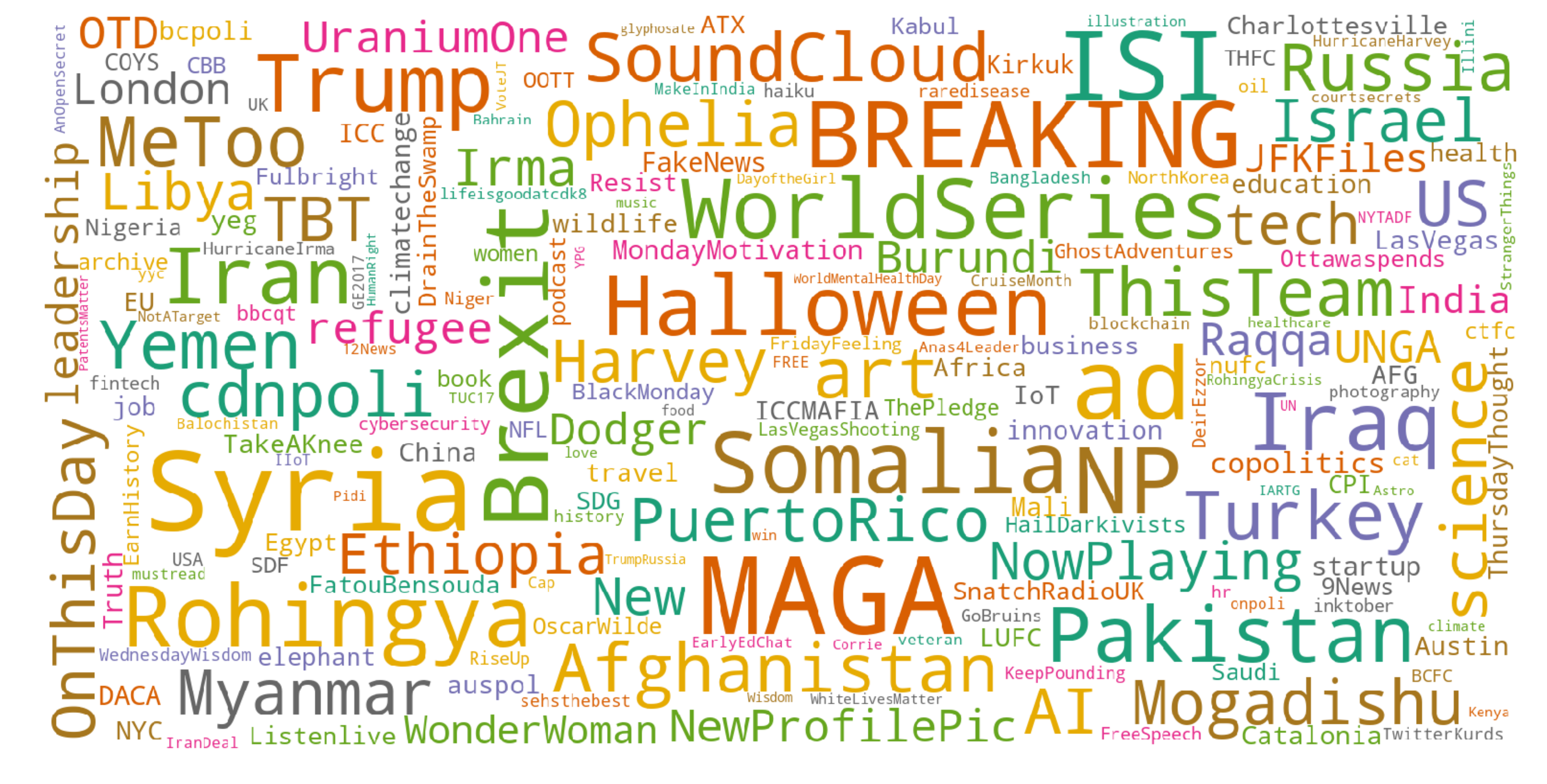}
        \caption{Word cloud from normal users' tweets. Notice that it shares several hashtags with the word cloud associated with hateful users, such as \texttt{MAGA} and \texttt{Syria}.}
        \label{fig:wc1}
    \end{minipage}\hspace{0.02\textwidth}
    \begin{minipage}{0.48\textwidth}
        \centering
        \includegraphics[width=0.96\linewidth]{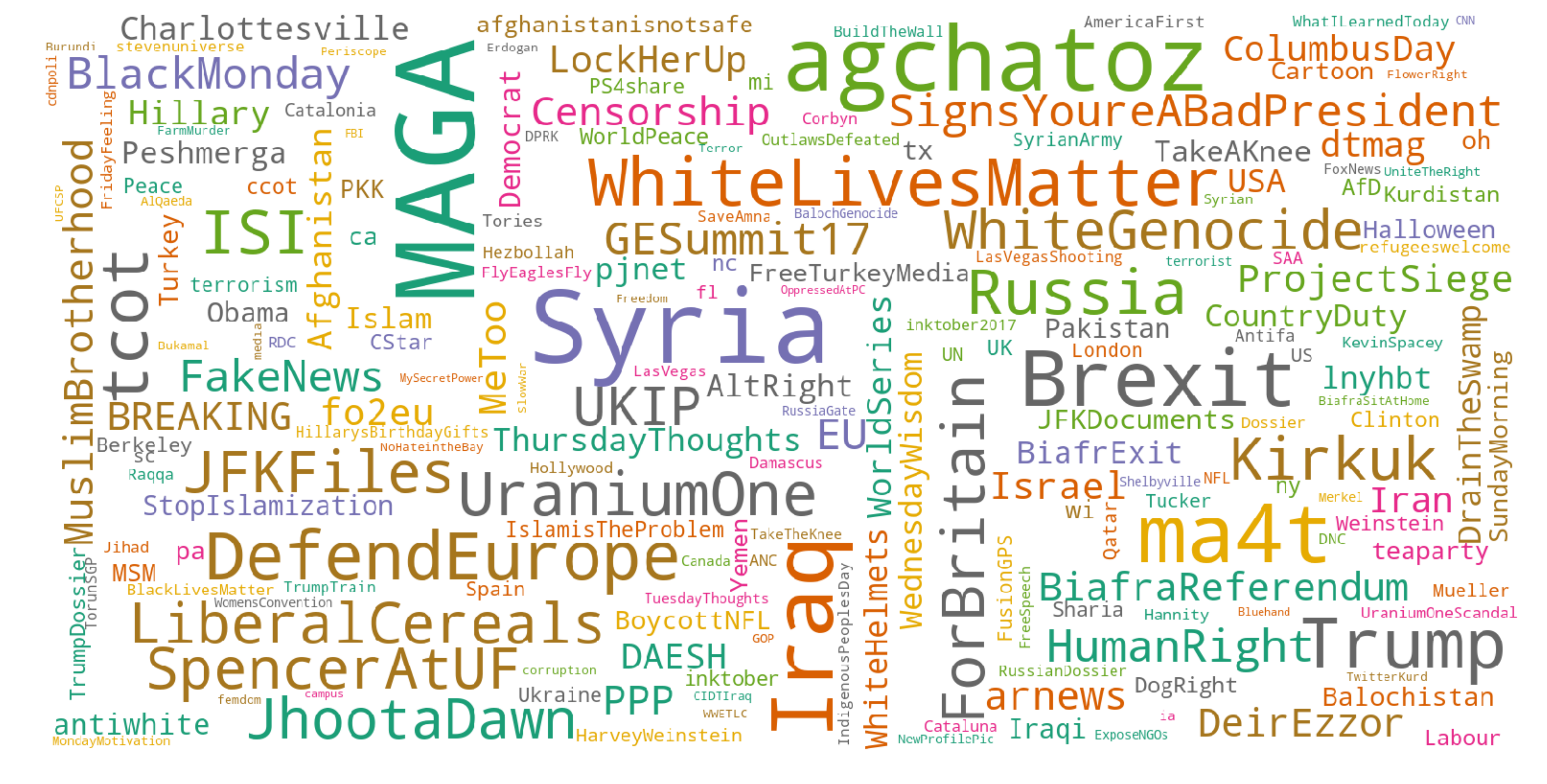}
        \caption{Word cloud from hateful users' tweets. Notice the inclusion of some hashtags associated with White Supremacist groups such as \texttt{WhiteGenocide}.}
        \label{fig:wc2}
    \end{minipage}
\end{figure*}
 
\paragraph{Lexical Analysis} We characterize hateful and normal users, as well as their neighborhood w.r.t. their content  with \textit{Empath}~\cite{fast2016empath}, as depicted in Figure~\ref{fig:lex}.
Our results are counter-intuitive. To begin with, hateful users use less words related to hate, anger, shame and terrorism, violence, and sadness (with p-values $< 0.001$), all of which are often taken as assumptions in the sampling process of other work intended to detect hateful tweets~\cite{kwok2013locate,davidson2017automated}. 
A question that rises in this context is how sampling tweets based exclusively in a hate-related lexicon biases the sample of content to be annotated to a very specific type of user, which may not be representative of the average "hate-spreading" ones.
This also reinforces the already stated claims that sarcasm and code-words may play a significantly role in defining such users~\cite{davidson2017automated, magu2017detecting}. 
Categories of words more used by hateful users include positive emotions, negative emotions, suffering, work, love and swearing (with p-values $< 0.001$). This suggests the use of emotional vocabulary by hateful users (and those in their 1-neighborhood). An interesting direction in that sense would be to analyze the sensationalism of statements made by hateful users when compared to normal ones, as it has been done in the context of \textit{clickbaits}, catchy titles often associated with frivolous or fake news-pieces~\cite{chen2015misleading}. 
Overall, the non-triviality of the lexical characteristics of these groups of users reinforces the difficulties found in the NLP community to attack the problem of successfully detecting hate-speech~\cite{davidson2017automated}.

\paragraph{Sentiment} Following on the finding that, according to \textit{Empath}, hateful users use more negative and positive words, we explore the sentiment in the sentences they write using  \textit{VADER}~\cite{yu2003towards}, as depicted in Figure~\ref{fig:sent}. We find  that sentences written by hateful users are more negative, and are less subjectivee (p-value $<0.001$). The neighborhood of hateful user is also more negative (p-value $<0.001$), however not less subjective. We also analyze the distribution profanity per tweet in hateful and non-hateful users. The latter is obtained by matching all the words in Shutterstock's "List of Dirty, Naughty, Obscene, and Otherwise Bad Words"~\footnote{https://github.com/LDNOOBW/List-of-Dirty-Naughty-Obscene-and-Otherwise-Bad-Words}. We find that hateful users and their neighborhood employ more profane words per tweet, also confirming the results from the analysis with \textit{Empath}. 

\begin{figure}[t]
\centering
        \includegraphics[width=0.8\linewidth]{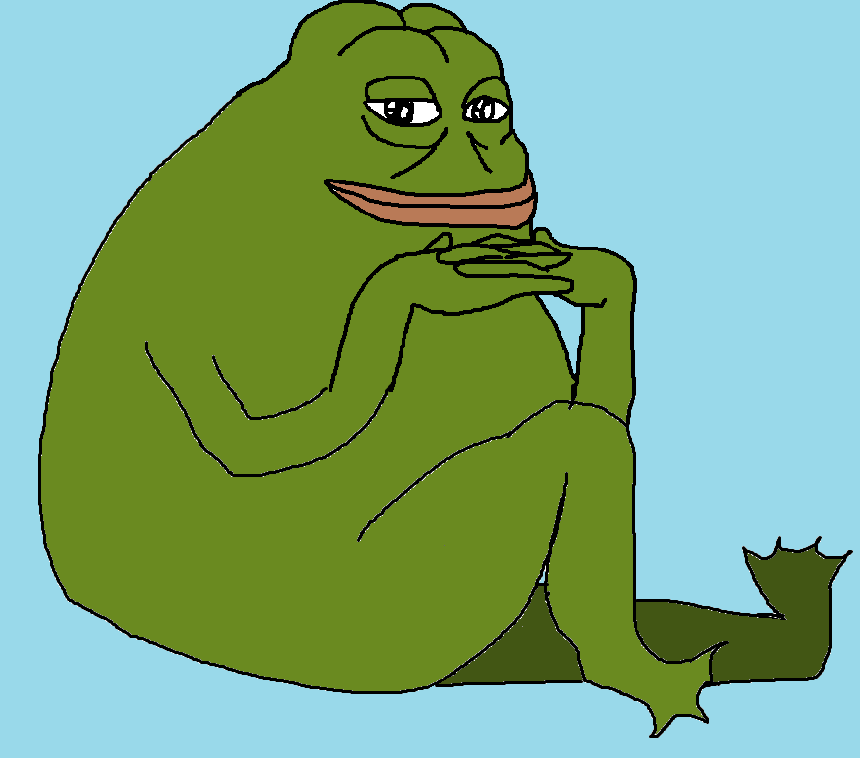}
        \caption{Groyper, an illustration of Pepe the Frog which was present in several hateful users identified, often in some variation.}
        \label{fig:groyper}
\end{figure}

\paragraph{A qualitative look} Finally, we briefly present two qualitative insights on the content present in the user profiles we analyze. In Figures~\ref{fig:wc1} and Figure~\ref{fig:wc2} we display wordclouds containing the hashtags that were mostly used by hateful and non-hateful users. The wordcloud for hateful users contains some hashtags that have been associated with openly racist institutions or individuals such as American Renaissance~\cite{noauthor_active_nodate}. Also, we can see that several hashtags are shared among both groups, such as \texttt{\#Iraq} or \texttt{\#MAGA}. Additionally, in Figure~\ref{fig:groyper} we show Groyper, a picture of Pepe the Frog resting on his chin, which originated in the imageboard 4chan, and is known commonly used as avatar among the alt-right and the new right in social media~\cite{noauthor_groyper_nodate}. An expressive number of the profiles identified as hateful by the annotators had Groyper (or some variation of Groyper) as a profile picture. These profiles are anonymous and tweet almost exclusively about politics, race and religion. Although we approach the problem of detecting hateful speech as a nuanced one, in the case of most of these profiles it is trivial to classify the vehiculated content according to the definition of hateful speech that we provided.

\paragraph{Suspended Accounts} Finally, we briefly analyze accounts that have been suspended three months after the data collection period in the $100$ thousand users we collect. Most Twitter accounts are suspended due to spam, however as these accounts rarely get retweeted, they are harder to reach in the retweet induced graph. Thus, we have that other common reasons for suspension are abusive behavior and security issues with the account. We find the accounts that have been suspended among the $100,386$ collected accounts by making requests to Twitter's API. We use these suspended accounts as another source for potentially hateful behavior, as quantitative and qualitative analysis suggests they do not behave as spammers, and as they have a large intersection with the accounts labeled as hateful. Notice that these accounts may present other types of abusive behavior other than hate speech, such as offenses not based on attributes such as race, gender, etc.

\begin{table}[h]
\centering
\caption{Percentage and absolute number of accounts that got suspended after three months}
\label{tab:sus}
\begin{tabular}{@{}llll@{}}
\toprule
 & Hateful   & Normal    & Others           \\ \midrule
Suspended Accounts                              & $$9.09\% (55)$$ & $$0.32\% (14)$$ & $$0.33\% (314)$$ \\ \bottomrule
\end{tabular}
\end{table}

\noindent
As depicted in Table~\ref{tab:sus}, we find that $55$ of the users classified as hateful by the crowdsourced annotators were banned in $3$ months time, which corresponds to roughly $9\%$ of all hateful users. In contrast, only $14$ normal users were banned ($0.32\%$), and for all $100$ thousand users, 314 users were banned, corresponding to $0.33\%$. This result strengthens our findings, as we find that the annotations we performed seem to be somewhat in accordance to Twitters own moderation process.
Interestingly, we collected the suspended accounts right before Twitter started to enforce new rules on violence, abuse, and hateful conduct, making exploring the differences between accounts that have been suspended \textit{before} and \textit{after} this change of policy a promising direction.
%
\newpage
\section{Related Work}
We briefly review previous work on detecting and characterizing hate speech in OSNs. Tangent problems such as cyber-bullying and offensive language are not extensively covered, refer to \cite{schmidt2017survey}. We compare aspects of other methodology previously employed. It is important to notice that, for many of the works done in the context of OSNs, the main objective of the work we refer was to detect hate speech, whereas we emphasize characterization.

Many previous studies collect data by sampling OSNs with the aid of a lexicon with terms associated with hate speech~\cite{davidson2017automated, waseem2016hateful, burnap2016us, magu2017detecting}. 
This may be succeeded by expanding this lexicon adding other co-occurring terms~\cite{waseem2016hateful}.
Other techniques employed include matching regular expressions~\cite{warner2012detecting}, selecting features in tweets from users known to have reproduced hate speech~\cite{kwok2013locate}.
We employ a random-walk-based methodology.
Unlike previous work, our methodology uses a lexicon of hate-related words as a starting point to run a diffusion process. This diffusion process will give us a number of "closeness to hate-related words" associated with each user, which we use to perform a stratified sampling of the users to be annotated.

In the existing previous work on hate-speech detection, human annotators are used to label content. 
This labeling may be done by the researchers themselves~\cite{waseem2016hateful, kwok2013locate,djuric2015hate, magu2017detecting}, 
selected annotators~\cite{warner2012detecting, gitari2015lexicon},
or crowd-sourcing services~\cite{burnap2016us}.
Hate-speech speech has been pointed out as a  difficult subject to annotate on~\cite{waseem2016you,ross2017measuring}.
We also employ \textit{CrowdFlower} to annotate our data. Unlike previous work we provide annotators with the the entire profile of the user instead of individual tweets, this provides better context for the annotators~\cite{waseem2017understanding}.

Although most previous works focus on detection, there are some notable exceptions. Silva et. al~\cite{silva2016analyzing}, matches regex-like expressions on large datasets on Twitter and Whisper to characterize the targets of hate in online social networks. Also, Gerstenfeld et. al~\cite{gerstenfeld2003hate} analyze hateful websites characterizing their \textit{modus operandi} w.r.t. monetization, recruitment, and international appeal.

%
\section{Discussion and Conclusion}
\label{sec:discussion}
We present a first characterization of hate speech in Online Social Networks at a user-level granularity. We develop a methodology to sample Twitter which consists of obtaining a generic subgraph in Twitter, finding users who employed words in a lexicon of hate-related words and running a diffusion process based on DeGroot's learning model to sample for users in the neighborhood of these users. We then used \textit{Crowdflower}  to manually annotate $4,972$ users, of which $544$ were considered to be hateful.

Our findings shed light on how hateful users are different from normal ones with respect to their user activity patterns, network centrality measurements, and the content they produce. Among our findings, we discover that the median hateful user is more central in the retweet network, more recently created, write more negative sentences and use lexicon associated with categories such as hate, terrorism, violence and anger \textit{less} than normal ones. Furthermore, this analysis seem to also hold for the 1-neighborhood of the hateful and normal users.

Nevertheless, our analysis still has limitations that lead to interesting future research directions. 
Firstly, it is reasonable to question the definition of \textit{hateful user}, in the sense that it is not clear what is the threshold an account has to violate to be considered hateful. 
Although we argue that classifying hateful users is easier than classifying hateful content, it is still a non-trivial task due to the subjectivity of the definition of hate-speech.
Secondly, it is not clear whether the characterization (and possibly detection) of hateful \textit{users} would solve all problems related to hate speech, as looking at this coarser-grained level of OSNs may make detecting users who only occasionally propagate hate speech harder. Thus, an interesting question in this scenario is \textit{How much of the hate speech is produced by what percentage of users?} Another weakness of our characterization is that we only considered the behavior of such users on Twitter, and it is possible that this analysis does not hold in other widely used OSNs, such as Facebook or Instagram.

As future work, we want to detection of hateful users OSNs, a task which may be explored in different ways. A simple strategy would be to develop classification models based on the numerical attributes that are associated with each user and analyzed in this paper, and with representations for the text employed by the users, such as \textit{word2vec}~\cite{mikolov2013efficient}. However, another exciting strategy would be to use the connections in the entire graph that we sampled in Twitter to create representations for each node (user). Interestingly, modern approaches allow each node to be linked to a vector of features~\cite{hamilton2017inductive}, which suggests that we would be able to use both the content produced by the user as well as their positions in the network.
If accomplished, such methods for detecting these misbehaving users could help the moderation teams of online social network to quickly identify and take the necessary measures against the hateful profiles.
%
\section*{Acknowledgements}
This is work was supported by CNPq, CAPES, FAPEMIG, InWeb,
MASWEB, INCT-Cyber, and ATMOSPHERE PROJECT. We would like to thank Nikki Bourassa, Ryan Budish, Amar Ashar and Robert Faris from the Berkman Klein Center for Internet and Society for their insightful suggestions.
%
\balance
\bibliographystyle{ACM-Reference-Format}
\bibliography{sigproc} 

\end{document}